\documentclass[
    amsmath,
    amsssymb,
    notitlepage,
    aps,
    ,superscriptaddress
    ]{revtex4-2} 
\pdfoutput=1
\usepackage{graphicx} \graphicspath{{./figs/}}
\usepackage{dcolumn}
\usepackage{bm}
\usepackage[version=4]{mhchem}
\usepackage{braket}
\usepackage{floatrow}
\usepackage[caption=false]{subfig}
\usepackage{gensymb}
\usepackage{threeparttable}
\usepackage{multirow}
\usepackage[usenames,dvipsnames]{color}
\usepackage{wasysym}
\usepackage{mathtools}
\usepackage[dvipsnames]{xcolor}
\usepackage[normalem]{ulem}
\usepackage[colorlinks]{hyperref}
\usepackage{hyperref} \hypersetup{colorlinks=true,citecolor=black,linkcolor=black,urlcolor=blue}
\usepackage{lineno}
\usepackage{lipsum} 

\makeatletter
\def\l@subsection#1#2{}
\def\l@subsubsection#1#2{}
\makeatother

    \makeatletter
\def\@fnsymbol#1{\ensuremath{\ifcase#1\or \dagger\or *\or \ddagger\or
   \mathsection\or \mathparagraph\or \|\or **\or \dagger\dagger
   \or \ddagger\ddagger \else\@ctrerr\fi}}
    \makeatother
\DeclareUnicodeCharacter{2212}{-}
\begin{document}
\title{Spin-polarized antiferromagnetic metals}
\author{Soho Shim}
\email{jsshim2@illinois.edu}
\affiliation{Department of Physics, University of Illinois at Urbana-Champaign, Urbana, IL 61801, USA}
\author{M. Mehraeen}
\affiliation{Department of Physics, Case Western Reserve University, Cleveland, OH 44106, USA}
\author{Joseph Sklenar}
\affiliation{Department of Physics and Astronomy, Wayne State University, Detroit, MI 48201, USA}
\author{Steven S.-L. Zhang}
\affiliation{Department of Physics, Case Western Reserve University, Cleveland, OH 44106, USA}
\author{Axel Hoffmann}
\affiliation{Department of Materials Science and Engineering, University of Illinois at Urbana-Champaign, Urbana, IL 61801, USA}
\author{Nadya Mason}
\email{nmason1@uchicago.edu}
\affiliation{Pritzker School of Molecular Engineering, University of Chicago, Chicago, IL 60637, USA}

\begin{abstract}

Spin-polarized antiferromagnets have recently gained significant interest because they combine the advantages of both ferromagnets (spin polarization) and antiferromagnets (absence of net magnetization) for spintronics applications. In particular, spin-polarized antiferromagnetic metals can be useful as active spintronics materials because of their high electrical and thermal conductivities and their ability to host strong interactions between charge transport and magnetic spin textures. We review spin and charge transport phenomena in spin-polarized antiferromagnetic metals, in which the interplay of metallic conductivity and spin-split bands offers novel practical applications and new fundamental insights into antiferromagnetism. We focus on three types of antiferromagnets: canted antiferromagnets, noncollinear antiferromagnets, and collinear altermagnets. We also discuss how the investigation of spin-polarized antiferromagnetic metals can open doors to future research directions.
\par \medskip
\small	
  \textbf{\textit{Keywords---}}antiferromagnetism, spin splitting, spin texture, charge transport, altermagnetism
\end{abstract}

\maketitle

\tableofcontents
\section{Introduction} \label{sec:introduction}

Spintronics, which utilizes the coupling between spin and charge degrees of freedom, has been an active research field due to its potential for non-volatile, energy-efficient data storage applications~\cite{dieny2020opportunities,puebla2020spintronic,hoffmann2015opportunities,hirohata2020review}. Ferromagnets have been central components to the development of spintronics because their spontaneous magnetization and the resulting momentum-independent spin-split bands allow the manipulation and detection of magnetic orders to be straightforward. However, spontaneous magnetization also limits the scalability of ferromagnetic spintronics devices.  Such limits motivate the study of antiferromagnets as alternative candidates for spintronics materials.

Antiferromagnets, due to their compensated magnetizations and lack of net magnetic moments, have many appealing properties as active spintronics materials~\cite{jungwirth2016antiferromagnetic,baltz2018antiferromagnetic}; they are robust against external magnetic fields, do not produce stray fields, and enable THz ultrafast dynamics. Several theoretical and experimental advances suggest that antiferromagnetic order can be manipulated and detected through various routes, such as Néel spin-orbit torque-based control for antiferromagnets with locally broken inversion symmetry~\cite{vzelezny2014relativistic,wadley2016electrical,bodnar2018writing}. However, conventional antiferromagnets, which only have two magnetic atoms with antiparallel moments and no other local symmetry breaking for the magnetic sublattices, have spin degenerate bands due to parity-time-reversal $\mathcal{PT}$ symmetry [the combination of parity ($\mathcal{P}$) and time-reversal symmetry ($\mathcal{T}$)] and do not have a corresponding spin polarization of the electronic band structure in momentum space (Fig.~\ref{fig:1_a}). Thus, the magnetic states of conventional antiferromagnets cannot be readily detected with typical macroscopic electrical or optical probes, and it is thus difficult to implement them in spintronics devices.

Recently, it has been demonstrated that certain antiferromagnets can possess spin-split bands when $\mathcal{PT}$ symmetry is broken~\cite{vsmejkal2022anomalous,cheong2024altermagnetism,vsmejkal2022emerging}, which lifts the Kramers' degeneracy (Figs.~\ref{fig:1_b}, \ref{fig:1_c}, and \ref{fig:1_d}). The spin-split bands in antiferromagnets thus enable leveraging the advantages of both ferromagnetism (spin-split bands) and antiferromagnetism (zero net magnetization).

Spin-polarized antiferromagnets have recently been investigated for various materials that break $\mathcal{PT}$ symmetry in different ways. For instance, noncollinear antiferromagnets, which break $\mathcal{PT}$ through noncollinear spin structures~(Fig.~\ref{fig:1_c}), have been observed to host the anomalous Hall effect~\cite{chen2014anomalous,kubler2014non,nakatsuji2015large,kiyohara2016giant,nayak2016large,tsai2020electrical,suzuki2017cluster,liu2018electrical,higo2018anomalous,taylor2020anomalous}.  This discovery initiated the search for other antiferromagnets that also host spin-split bands. Canted antiferromagnets, which break $\mathcal{PT}$ through the presence of a net magnetization~(Fig.~\ref{fig:1_b}), are also reported to host spin-split bands~\cite{vsmejkal2022anomalous,kipp2021chiral,cao2023plane} and an associated spontaneous Hall effect~\cite{kipp2021chiral,ibarra2022noncollinear,ibarra2022anomalous}. Beyond noncollinear antiferromagnets, collinear ``altermagnets", whose alternating local crystal structures around spin sites break $\mathcal{PT}$ symmetry~(Fig.~\ref{fig:1_d}), have also been reported to host spin-split bands and similar emergent phenomena~\cite{vsmejkal2022emerging,feng2022anomalous}.

As different types of spin-polarized antiferromagnets have recently been identified, there have been increasing efforts to develop a holistic understanding of these materials. For instance, while the term ``altermagnetism" initially only referred to collinear antiferromagnets~\cite{vsmejkal2022beyond,vsmejkal2022emerging}, recent works have proposed that the definition of altermagnetism can be extended to general antiferromagnets with momentum-dependent spin-split bands, such as noncollinear antiferromagnets~\cite{zhu2024observation,cheong2024altermagnetism}. In addition to ongoing efforts on a fundamental understanding of these systems, many key spintronics phenomena in spin-polarized antiferromagnets have been experimentally demonstrated, such as room-temperature all-antiferromagnetic tunnel junctions based on a noncollinear antiferromagnet~\cite{qin2023room,chen2023octupole}, and spin current generation via magnetic spin Hall effect~\cite{bose2022tilted,bai2022observation} and electrical 180\degree\ switching of the Néel vector~\cite{han2024electrical} in a collinear altermagnet. While still in the early stages, the investigation of spin-polarized antiferromagnets is rapidly evolving with new theoretical and experimental results.

Detailed reviews focusing on various aspects of spin-polarized antiferromagnets have been published elsewhere~\cite{vsmejkal2022anomalous,chen2024emerging,yan2024review,cheong2024altermagnetism}. In this review, we discuss how different charge and spin transport phenomena can arise in spin-polarized antiferromagnets. We also focus on antiferromagnetic metals, in which the spin polarization of the conduction electrons enables direct coupling between charge transport and magnetic spin textures~\cite{siddiqui2020metallic}. We first introduce and discuss the key properties of different spin-polarized antiferromagnets based on the manner in which $\mathcal{PT}$ symmetry is broken (Sec.~\ref{sec:magnetic}). These systems include canted antiferromagnets~(Fig.~\ref{fig:1_b}), noncollinear antiferromagnets~(Fig.~\ref{fig:1_c}), and collinear altermagnets~(Fig.~\ref{fig:1_d}). We then discuss how spin-polarized antiferromagnetic metals can host different charge transport phenomena. We discuss how spin-split bands in spin-polarized antiferromagnetic metals enable the readout of antiferromagnetic states via linear magnetotransport effects (Sec.~\ref{sec:linear}: magnetoresistance, anomalous Hall effect) and nonlinear magnetotransport effects (Sec.~\ref{sec:nonlinear}: unidirectional magnetoresistance, nonlinear Hall effect). We then discuss how spin-polarized antiferromagnetic metals allow the investigation of spin texture topology via nonlinear Hall effect measurements (Sec.~\ref{sec:nonlinear_NLHE}). Finally, we discuss how spin-polarized antiferromagnetic metals can generate spin currents based on their spin-split bands (Sec.~\ref{sec:spin current}) as well as generate spin torques (Sec.~\ref{sec:spintorques}).

\section{Magnetic orders} \label{sec:magnetic}
\begin{figure}[hbt!]
    \sidesubfloat[]{\includegraphics[width=0.8\linewidth]{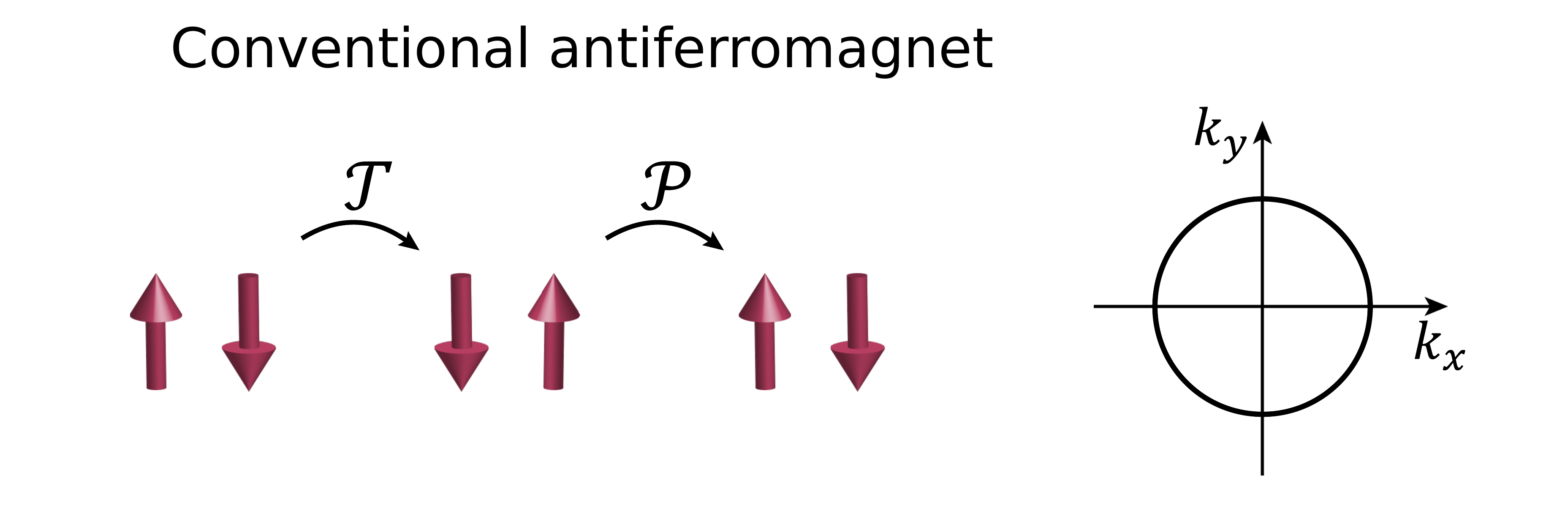}\label{fig:1_a}}\quad \quad%
    \sidesubfloat[]{\includegraphics[width=0.8\linewidth]{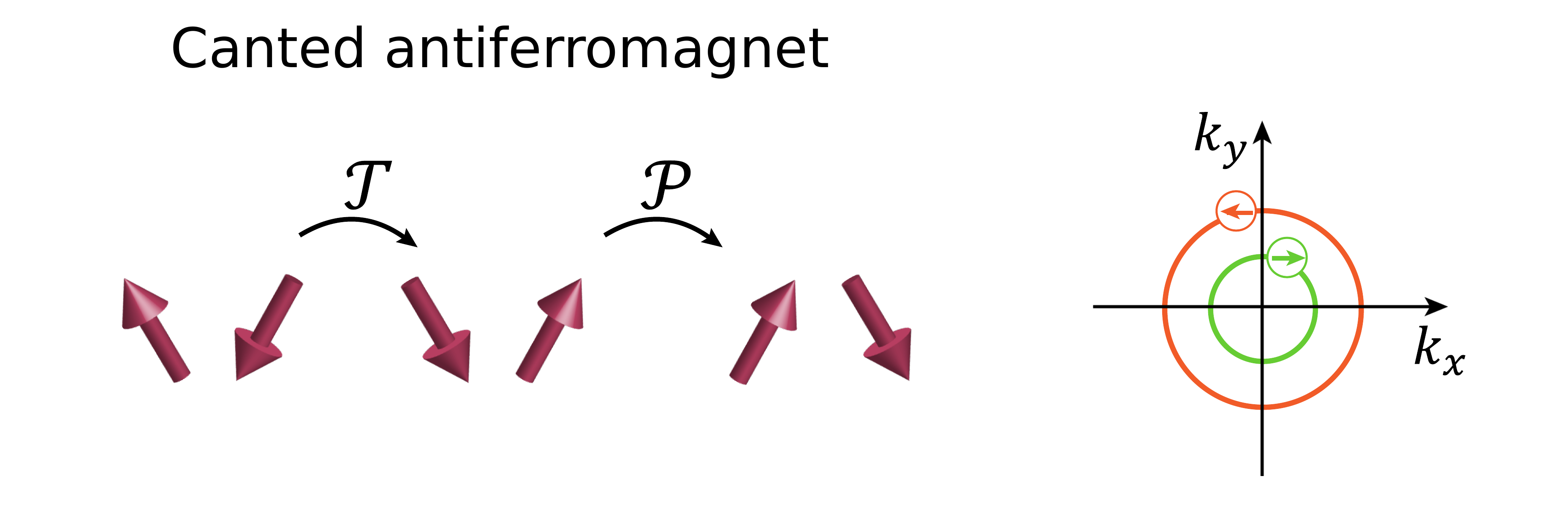}\label{fig:1_b}}\quad%
    \sidesubfloat[]{\includegraphics[width=0.8\linewidth]{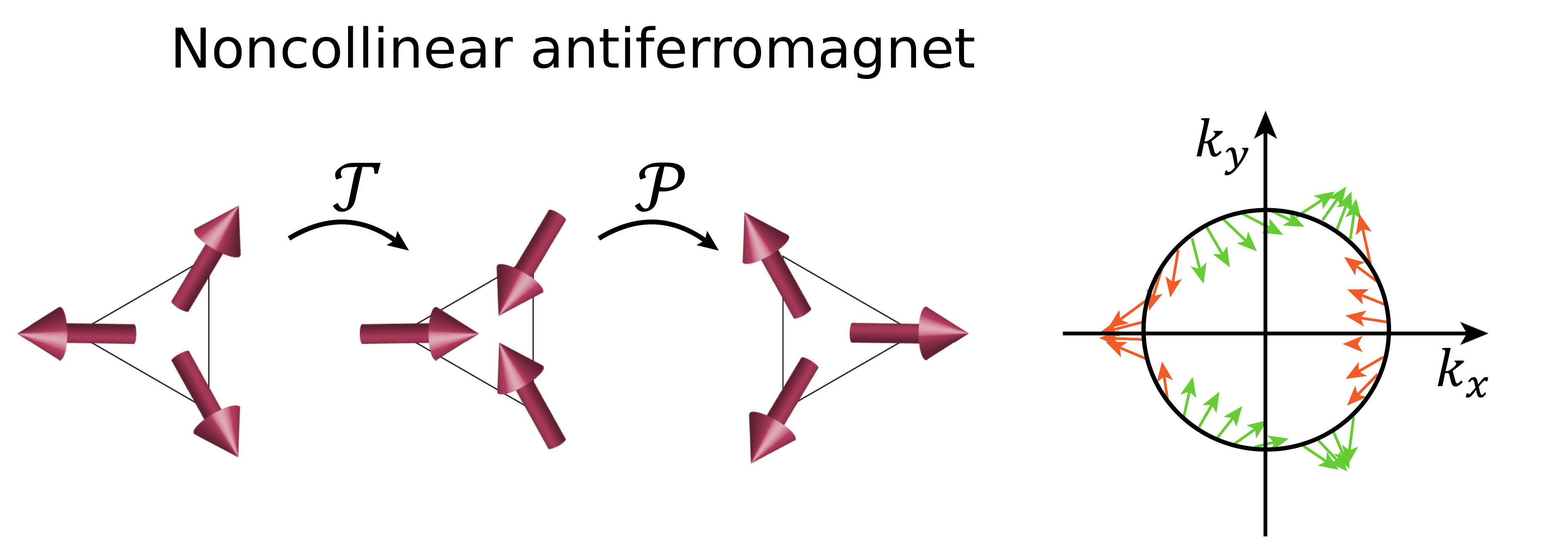}\label{fig:1_c}}\quad%
    \sidesubfloat[]{\includegraphics[width=0.8\linewidth]{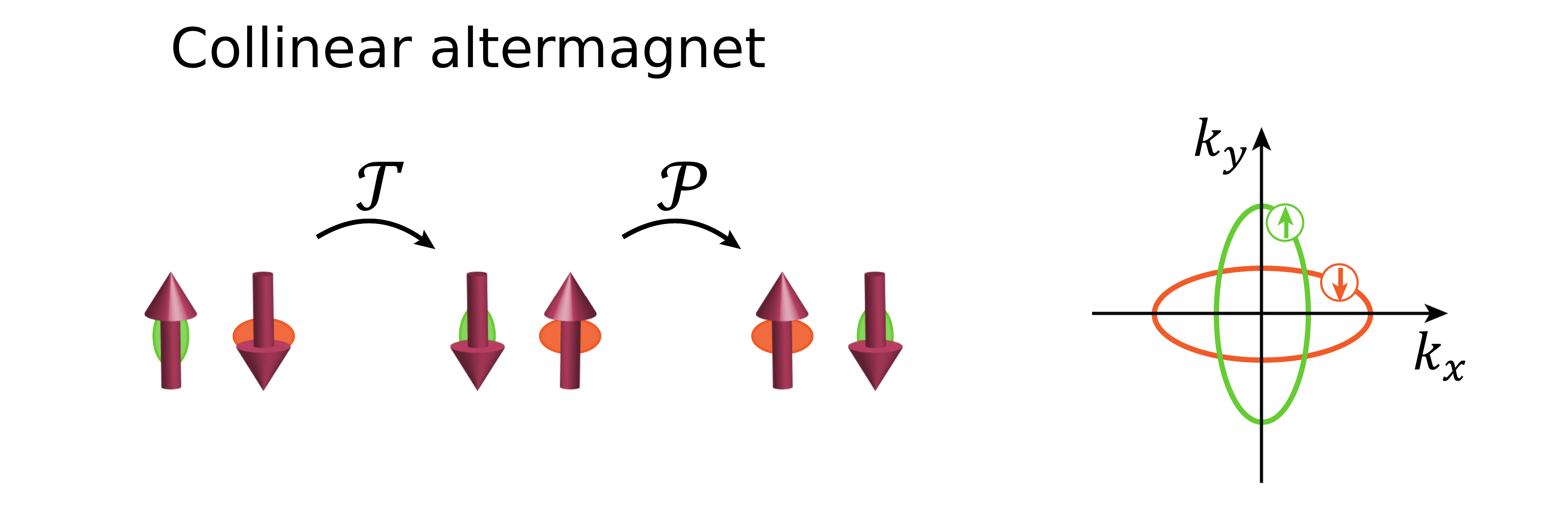}\label{fig:1_d}}%
    \caption{Symmetry properties and corresponding Fermi surface of different antiferromagnetic orders: \normalfont{(a) $\mathcal{PT}$ symmetry is preserved in conventional antiferromagnets, such as in MnPt or IrPt, thus the bands are spin-degenerate and the Fermi surfaces are centrosymmetric. (b) Conventional collinear antiferromagnets can have spin canting due to either Dzyaloshinskii-Moriya interaction (DMI) or external magnetic field. The canted spin structure breaks $\mathcal{PT}$ symmetry, which leads to spin-split bands in canted antiferromagnets, such as in GdPtBi. Green/orange arrows and lines in the Fermi surfaces indicate spin polarization and spin-split bands respectively. (c) The noncollinear spin structure breaks $\mathcal{PT}$ symmetry, which induces momentum-dependent spin-split bands in noncollinear antiferromagnets. The Fermi surface shows the spin texture in momentum space for noncollinear coplanar antiferromagnets on a kagomé lattice ($\Gamma_{5g}$ magnetic configuration), such as in Mn$_3$Ir, Mn$_3$Rh, and Mn$_3$Pt. The Fermi surface diagram is inspired by and adapted with permission from Ref.~\cite{vzelezny2017spin}. Copyrighted by the American Physical Society. (d) Alternating local crystal structures (green and orange ellipsoids) around spin sites break $\mathcal{PT}$ symmetry and lead to momentum-dependent spin-split bands and anisotropic spin-split Fermi surfaces in collinear altermagnets, such as in RuO$_2$. The Fermi surface diagram is inspired by Ref.~\cite{vsmejkal2022emerging} (CC 4.0 International)}}\label{fig:1}
\end{figure}

For antiferromagnetic metals to host strong spin-polarization phenomena similar to those in ferromagnetic metals, it necessitates introducing spin-split band structures. For conventional antiferromagnets, the combination of $\mathcal{P}$ and $\mathcal{T}$ is preserved whereby all the bulk bands are spin-degenerate at all momenta in the Brillouin zone~(Fig.~\ref{fig:1_a})~\cite{vsmejkal2022anomalous,vsmejkal2022emerging,cheong2024emergent}. The Kramers' degeneracy may be lifted in antiferromagnets with $\mathcal{PT}$ symmetry breaking, allowing the emergence of spin-split bands~(Figs.~\ref{fig:1_b}, ~\ref{fig:1_c}, and ~\ref{fig:1_d})~\cite{cheong2024altermagnetism}.

In the following subsections, we introduce three different types of spin-polarized antiferromagnets that host spin-split bands in momentum space due to broken $\mathcal{PT}$ symmetry: A) canted antiferromagnets  (Fig.~\ref{fig:1_b}), B) noncollinear antiferromagnets (Fig.~\ref{fig:1_c}), and C) collinear altermagnets (Fig.~\ref{fig:1_d}). More detailed discussions of the classification of these emerging magnetic materials based on symmetry considerations and group theory can be found in Refs.~\cite{winkler2023theory, vsmejkal2022anomalous, vsmejkal2022emerging, cheong2024altermagnetism}.

\subsection{Canted antiferromagnets} \label{sec:magnetic_canted}
Spin canting in collinear antiferromagnets can result in spin-polarized bands~\cite{cao2023plane,yan2024review} (Fig.~\ref{fig:1_b}). The canting can be induced by an external magnetic field or thermal fluctuations, and it may also arise intrinsically from the Dzyaloshinskii-Moriya interaction (DMI)---an antisymmetric exchange interaction due to spin-orbit coupling and inversion symmetry breaking. In the latter case, the DMI couples the net magnetization $\mathbf{M}$ and the Néel vector $\mathbf{L}$, giving rise to weak ferromagnetism (ferrimagnetism) with spin-split bands~\cite{vsmejkal2022anomalous}. The spin canting can be quantified by the vector spin chirality $\mathbf{S}_{i} \times \mathbf{S}_{j}$, which may be linked with the momentum-space Berry curvature of the spin-split bands~\cite{kipp2021chiral}.

Unlike conventional spin-polarized conducting ferromagnets, emergent transport properties in canted antiferromagnets ({\em e.g.}, the anomalous Hall effect) may be driven primarily by the Néel vector $\mathbf{L}$ rather than the small net magnetization $\mathbf{M}$~\cite{vsmejkal2020crystal,feng2022anomalous}, given the interplay between the large antiferromagnetic exchange coupling and the weak ferromagnetism. Canted antiferromagnetic metals that have received considerable attention include MnPtGa~\cite{ibarra2022noncollinear,you2019competitive,cooley2020evolution,ibarra2022anomalous} and CuMnSb~\cite{regnat2018canted}, wherein intrinsic spin canting is induced by competing exchange interactions, and a family of half-Heusler compounds RPtBi (R = Gd, Nd, Tb, Dy)~\cite{suzuki2016large,shekhar2018anomalous,zhu2020exceptionally,zhang2020field}, some of which are regarded as magnetic Weyl semimetals wherein the strong correlation between magnetic structure and topologically nontrivial electronic band structures near the Fermi surface may give rise to a number of emergent magnetotransport effects~\cite{suzuki2016large,shekhar2018anomalous,li2021nonlinear}.\\

\subsection{Noncollinear antiferromagnets} \label{sec:magnetic_noncollinear}

A noncollinear antiferromagnet exhibits a noncollinear magnetic structure typically sitting on a triangular or Kagomé lattice with antiferromagnetic coupling between local magnetic moments that may arrange themselves into a multiplicity of degenerate ground states due to geometric frustration. In contrast to collinear antiferromagnets (whose magnetic order can be defined by a single Néel vector), the magnetic order of noncollinear antiferromagnets cannot be represented by a single vector. As a result, there is no well-defined spin-quantization axis for conduction electrons in noncollinear antiferromagnetic metals. Furthermore, noncollinear antiferromagnets with broken $\mathcal{PT}$ symmetry may host spin-split bands with momentum-dependent spin states (Fig.~\ref{fig:1_c}), giving rise to intriguing spin-dependent transport phenomena that can be exploited for electrical characterization of the antiferromagnetic order~\cite{chen2014anomalous,kubler2014non,nakatsuji2015large,kiyohara2016giant,nayak2016large,tsai2020electrical,suzuki2017cluster,liu2018electrical,higo2018anomalous,taylor2020anomalous} and magnetically-controllable spin current generation~\cite{kimata2019magnetic,holanda2020magnetic}.

A widely studied class of noncollinear antiferromagnetic metals are Mn$_{3}$X (X = Ir, Sn, Ge, Pt)~\cite{chen2014anomalous,kubler2014non,nakatsuji2015large,kiyohara2016giant,nayak2016large,tsai2020electrical,suzuki2017cluster,liu2018electrical,higo2018anomalous,taylor2020anomalous}, for which a magnetic octupole order can be defined~\cite{suzuki2017cluster,suzuki2019multipole}. In addition, the application of an external magnetic field to these (coplanar) noncollinear  antiferromagnets~\cite{li2023field} or a slight distortion of their lattice structures~\cite{gong2016emergent,domenge2005twelve,shindou2001orbital} would induce an out-of-plane spin canting, giving rise to real-space spin chirality [which can be described by $\chi = \mathbf{S}_{i}\cdot (\mathbf{S}_{j} \times \mathbf{S}_{k})$---a scalar that is also adopted to characterize the real-space topological spin texture of magnetic skyrmions]. Several novel transport effects arising from the interplay between the real- and momentum-space spin textures in these materials have been reported~\cite{li2023field, gong2016emergent,domenge2005twelve,shindou2001orbital}.  

\subsection{Collinear altermagnets} \label{sec:magnetic_collinear}

Collinear altermagnets are a new class of magnetic materials, which do not fall under the dichotomy of ferromagnetism and antiferromagnetism~\cite{vsmejkal2020crystal,hayami2019momentum,yuan2020giant,mazin2021prediction,vsmejkal2022emerging}. In real space, they exhibit both an alternating magnetic structure (just as collinear antiferromagnets) and an alternating local crystalline structure. Such a doubly alternating pattern breaks $\mathcal{PT}$ symmetry and makes collinear altermagnets distinctly different from their antiferromagnetic and ferromagnetic counterparts (Fig.~\ref{fig:1_d}). As compared to collinear antiferromagnets (whose sublattices with opposite spins are related by either inversion or lattice translation), the opposite-spin sublattices of altermagnets are related by rotation. In other words, collinear altermagnets and conventional antiferromagnets have different spin-group symmetries~\cite{vsmejkal2022beyond}. These differences in real space are also reflected in the spin characters of their band structures: for conventional collinear antiferromagnets, the bands are spin-degenerate [i.e., $\varepsilon(s, \mathbf{k})=\varepsilon(-s, \mathbf{k})$], whereas band dispersions of collinear altermagnets show characteristics of both $\mathcal{T}$-symmetry breaking [i.e., $\varepsilon(s, \mathbf{k})\neq\varepsilon(-s, -\mathbf{k})$] and spin-splitting [i.e., $\varepsilon(s, \mathbf{k})\neq\varepsilon(-s,\mathbf{k})$] even without spin-orbit coupling. It is worth noting that for collinear altermagnets, spin is a good quantum number as one can still define a momentum-independent spin quantization axis that is collinear with the Neel vector. And as compared to (collinear) ferromagnetic metals, which also lack $\mathcal{PT}$-symmetry and hence possess spin-split bands and Fermi surfaces, altermagnets have anisotropic $d$ ($g$- or $i$-)-wave Fermi surfaces~\cite{vsmejkal2022emerging,bhowal2024ferroically} (Fig.~\ref{fig:1_d}), as opposed to isotropic $s$-wave ones.  

Altermagnetism has been predicted to emerge in the full span of conduction types, ranging from superconductors to insulators, and several systems have been identified as altermagnets based on direct observation of band splitting with angle-resolved photoemission spectroscopy~\cite{fedchenko2024observation,reimers2024direct,osumi2024observation,zhu2024observation}. Notable examples of metallic altermagnets that are identified or proposed are RuO$_2$~\cite{ahn2019antiferromagnetism, vsmejkal2020crystal,fedchenko2024observation}, FeSb$_2$~\cite{mazin2021prediction, vsmejkal2022beyond}, CrSb~\cite{vsmejkal2022beyond,reimers2024direct}, VNb$_3$S$_6$~\cite{vsmejkal2022beyond} and CoNb$_3$S$_6$~\cite{vsmejkal2020crystal}. A variety of emerging magnetotransport phenomena have been observed in these materials, arising from their distinct magnetic and electronic band structures. 

\section{Linear magnetotransport} \label{sec:linear}
\begin{figure}[hbt!]
    \sidesubfloat[]{\includegraphics[width=0.3\linewidth]{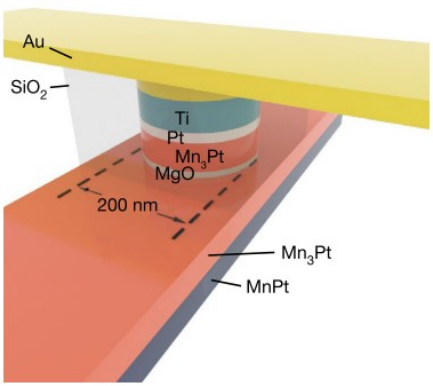}\label{fig:2_a}}\quad \quad%
    \sidesubfloat[]{\includegraphics[width=0.3\linewidth]{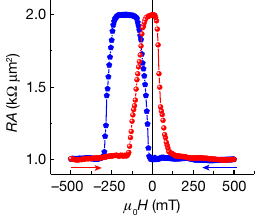}\label{fig:2_b}}\quad%
    \sidesubfloat[]{\includegraphics[width=0.25\linewidth]{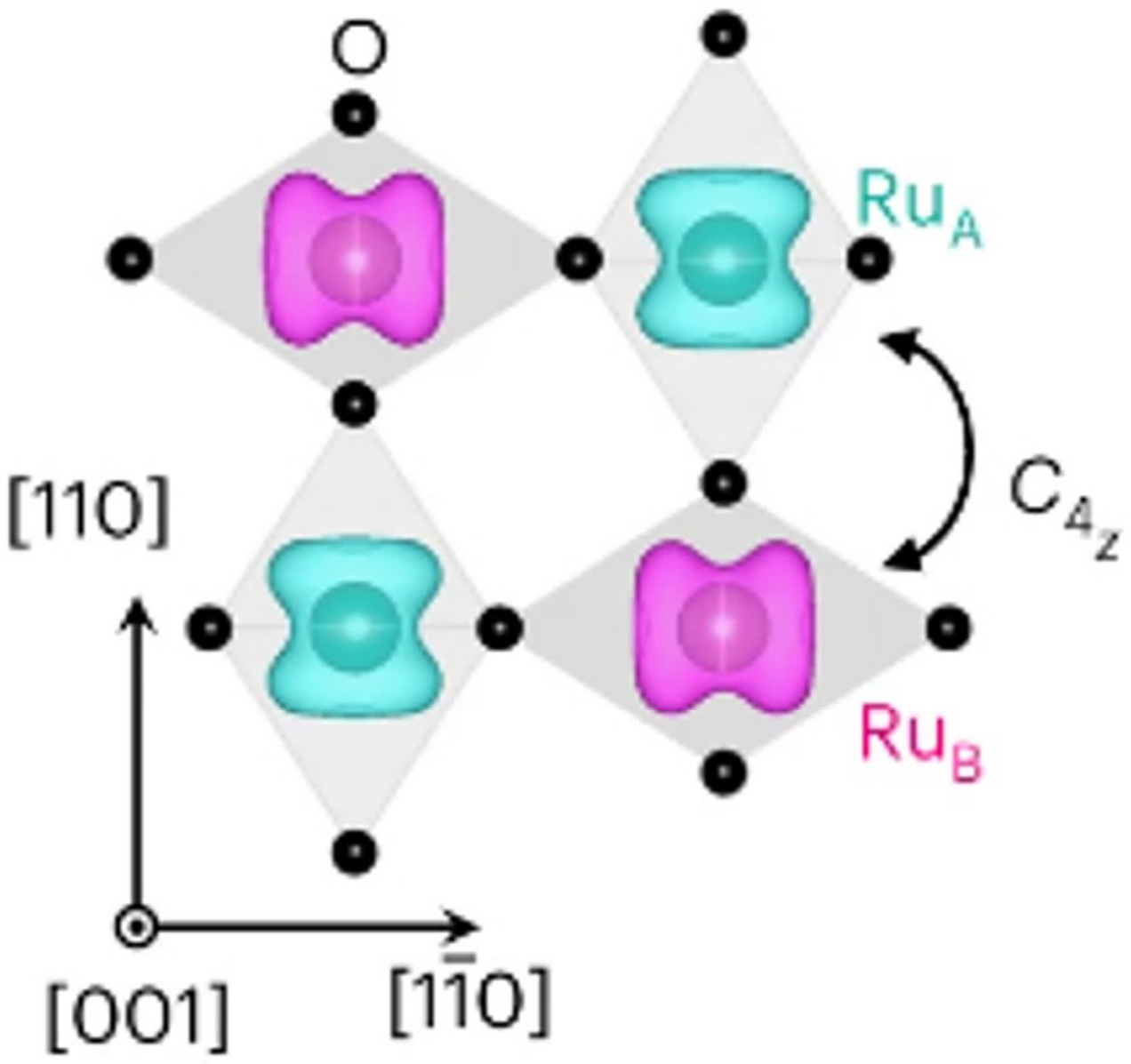}\label{fig:2_c}}\quad%
    \sidesubfloat[]{\includegraphics[width=0.3\linewidth]{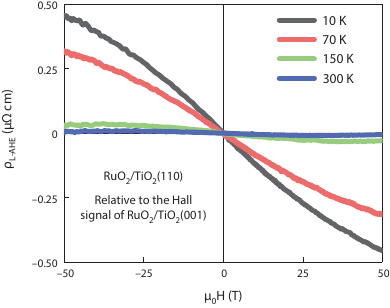}\label{fig:2_d}}%
\caption{Linear magnetotransport phenomena in spin-polarized antiferromagnets: \normalfont{(a,b) All-antiferromagnetic tunnel junction (AATJ) devices based on noncollinear antiferromagnet Mn$_3$Pt. (a) Schematic of a tunnel junction device with Mn$_3$Pt as key functional layers. A stack of Au/Ti/Pt multilayers is used as the top electrode. (b) Room temperature tunneling magnetoresistance (TMR) with the out-of-plane magnetic field sweep in a representative AATJ device. Panels (a) and (b) reprinted with permission from Ref.~\cite{qin2023room}, Springer Nature. (c,d) Observation of anomalous Hall effect in collinear altermagnet RuO$_2$. (c) Schematic of the RuO$_2$ crystal with opposite spin sublattices and ab initio spin density isosurfaces of Ru atoms expressed as green and magenta colours. The O atoms are depicted by the black dots. The black double arrow and its label illustrate a fourfold real-space rotation symmetry in RuO$_2$. (d) Anomalous Hall resistivity due to the symmetry breaking by the altermagnetic order ($\rho_{L-AHE}$). Panels (c) and (d) reprinted with permission from Ref.~\cite{feng2022anomalous}, Springer Nature.}}\label{fig:2}
\end{figure}

Linear magnetotransport phenomena, including magnetoresistance and the anomalous Hall effect, are widely utilized for sensing the presence or orientation of magnetic order. Magnetoresistance has been implemented for sensing the orientation of magnetic order and the anomalous Hall effect has served as a macroscopic probe for time-reversal symmetry breaking in magnetic systems. Due to the absence of spin polarization, conventional antiferromagnets are either invisible to many linear magnetotransport probes ({\em e.g.}, anomalous Hall effect, giant magnetoresistance, and tunneling magnetoresistance) or their readout signal based on a few magnetoresistance probes [{\em e.g.}, anisotropic magnetoresistance (AMR)] is small in magnitude. In this section, we discuss how spin-polarized conduction electrons in metallic antiferromagnets with broken $\mathcal{PT}$ enable the successful readout of antiferromagnetic orders and spin textures via magnetoresistance and the anomalous Hall effect.

\subsection{Magnetoresistance effects} \label{sec:linear_magnetoresistance}
Magnetoresistance is an extensively studied electrical property of magnetic metals where resistivity changes with the magnetization~\cite{mcguire1975anisotropic}; it is often utilized to sense the orientation of the magnetic order. For instance, AMR is used to read out the memory states of metallic antiferromagnets~\cite{wadley2016electrical,marti2014room,bodnar2018writing}.

Recent studies indicate that, in addition to probing the orientation of the antiferromagnetic order, AMR can also provide direct information about the spin textures near the Fermi level in antiferromagnets, especially ones with field-induced canting. For instance, a sign reversal of AMR from positive to negative with increasing in-plane magnetic field was recently reported for the metallic antiferromagnet FeRh~\cite{sklenar2023evidence}. This sign change was attributed to the concerted actions of field-induced spin canting and spin-orbit interaction and the resulting distortion of the Weyl Fermi surfaces. Similar sign changes in the AMR have been observed in other antiferromagnets with Weyl points as well, such as EuTiO$_3$~\cite{ahadi2019anisotropic} and Sr$_2$IrO$_4$~\cite{wang2014anisotropic}, and the sign reversal is similarly attributed to field-induced spin canting altering the Weyl points near the Fermi surface.

Beyond the AMR effect, recent giant magnetoresistance (GMR) or tunneling magnetoresistance (TMR) studies on spin-polarized antiferromagnets demonstrate the prospects of spin-polarized antiferromagnets as active spintronics components in magnetic junctions. Several groups have demonstrated proof-of-principle room temperature, all-antiferromagnetic tunnel junctions with the noncollinear antiferromagnets Mn$_3$Pt~\cite{qin2023room} (Figs.~\ref{fig:2_a} and ~\ref{fig:2_b}) and Mn$_3$Sn~\cite{chen2023octupole}. For instance, all-antiferromagnetic tunnel junctions based on the noncollinear antiferromagnet Mn$_3$Pt utilized the exchange bias effect between a collinear antiferromagnet, MnPt, and a noncollinear antiferromagnet, Mn$_3$Pt, which arises from the interfacial exchange coupling between the uncompensated spins in MnPt and the spins in Mn$_3$Pt~\cite{qin2023room}. This exchange-biased antiferromagnetic bilayer stack was fabricated into a tunnel junction device by depositing an insulating MgO layer and a thin Mn$_3$Pt capping layer which serves as a free noncollinear layer (Fig.~\ref{fig:2_a}). For a nominal 200-nm-diameter all-antiferromagnetic tunnel junction, room-temperature TMR reached approximately 100\% (Fig.~\ref{fig:2_b}), which is orders of magnitude larger than the signal of conventional single-layer antiferromagnetic spintronics devices based on the AMR detection~\cite{marti2014room,wadley2016electrical,bodnar2020magnetoresistance}. While the pinning of the bottom Mn$_3$Pt layer is based on the interfacial exchange coupling between MnPt and Mn$_3$Pt, first-principles calculations showed that the TMR effect originates from the momentum-dependent spin splitting in the noncollinear antiferromagnet Mn$_3$Pt and that a large TMR is expected for generic Mn$_3$Pt/barrier/Mn$_3$Pt junctions. Similar to Mn$_3$Pt tunnel junctions, the TMR effect in Mn$_3$Sn tunnel junctions is also based on the detection of the higher-order magnetic order which originates from the momentum-space Berry curvature in Mn$_3$Sn~\cite{chen2023octupole}. In addition to noncollinear antiferromagnets, collinear altermagnets are also predicted to host large GMR and TMR effects, arising from their spin-split band structures~\cite{vsmejkal2022giant,jiang2023prediction}. However, this has not yet been reported experimentally. Given that all-antiferromagnetic junctions have nearly vanishing stray fields and can host strongly enhanced spin dynamics up to the terahertz level, GMR and TMR studies on spin-polarized antiferromagnets hold great promise for the development of next-generation magnetic memory devices.

\subsection{Hall effects} \label{sec:linear_Hall}
The anomalous Hall effect (AHE), in which a transverse current flows due to spontaneous magnetization, can serve as a macroscopic probe of broken time-reversal symmetry in magnetic systems. While it was previously unexpected in antiferromagnets given their absence of net magnetization, spin-polarized antiferromagnets can generate large AHEs based on nonzero spin polarization.

Studies of the AHE in spin-polarized antiferromagnets have gained significant interest, particularly due to theoretical predictions~\cite{chen2014anomalous,kubler2014non} and its observation in noncollinear antiferromagnets at room temperature, namely in the  Mn$_{3}$X (X = Ir, Sn, Ge, Pt) family~\cite{chen2014anomalous,kubler2014non,nakatsuji2015large,kiyohara2016giant,nayak2016large,tsai2020electrical,suzuki2017cluster,liu2018electrical,higo2018anomalous,taylor2020anomalous}. The sizable AHE in noncollinear antiferromagnets has demonstrated how the Hall effect is closely related to topology and spin textures in the band structure~\cite{suzuki2017cluster,suzuki2019multipole}, thereby motivating its investigation in other antiferromagnets with spin-split bands.

The AHE can also arise when spin chirality in canted antiferromagnets induces a Berry curvature. In canted antiferromagnets, a momentum-space Berry curvature due to spin canting (equivalently, vector spin chirality) induces a chiral Hall effect, which scales linearly with the canting angle. The chiral Hall effect was proposed theoretically in \cite{kipp2021chiral} and it can be distinguished from other Hall effects ({\em e.g.}, crystal Hall effect and topological Hall effect) based on its linear scaling with canting angle or its energy scale. There have been several experimental reports of the chiral Hall effect in canted antiferromagnets, which demonstrate the role of spin canting and the associated vector spin chirality in inducing such linear magnetotransport phenomena. For instance, the AHE was reported for the metal MnPtGa in its canted antiferromagnetic state~\cite{ibarra2022noncollinear,ibarra2022anomalous}. The emergence of the chiral Hall effect in field-induced canted antiferromagnets is also reported for GdPtBi (a Weyl semimetal half-Heusler antiferromagnet)~\cite{suzuki2016large,shekhar2018anomalous} and FeRh (antiferromagnetic metal)~\cite{sklenar2023evidence}, in which the AHE contribution is attributed to field-induced canting. While a more thorough investigation is needed to distinguish the chiral Hall effect from other Hall effect contributions, such experimental reports of chiral Hall effects in canted antiferromagnets help explain the role of spin canting in complex magnets, and enable improved detection and control of various chiral magnetic states.

Collinear altermagnets can also host a strong AHE based on their anisotropic Berry curvatures in momentum space and broken $\mathcal{PT}$ symmetry. The AHE in collinear altermagnets is dubbed the ``crystal Hall effect", because it is sensitive to the crystal field created by the arrangement of non-magnetic atoms and it reverses its sign when reversing the arrangement of non-magnetic atoms between two magnetic sublattices (Fig.~\ref{fig:2_c}).

Following its prediction in collinear altermagnets~\cite{vsmejkal2020crystal}, a robust AHE has recently been reported in the metallic antiferromagnet RuO$_2$, with a magnitude comparable to ferromagnetic systems \cite{feng2022anomalous} (Figs.~\ref{fig:2_c} and ~\ref{fig:2_d}). In RuO$_2$, the broken $\mathcal{PT}$ symmetry arises from the crystal field created by the anisotropic arrangement of non-magnetic oxygen atoms. Instead of inversion symmetry, the two crystal sublattices with opposite Ru magnetic moments are connected by a crystal rotation symmetry when the non-magnetic O atoms are considered (Fig.~\ref{fig:2_c}). Such a rotational symmetry for opposite spin sublattices ensures the compensation of magnetizations but still allows broken $\mathcal{PT}$ and the resulting anisotropic Berry curvatures in momentum space. Fig.~\ref{fig:2_d} demonstrates the AHE contribution from antiferromagnetic order (L-AHE). Consistent with the theory prediction~\cite{vsmejkal2020crystal}, the L-AHE contribution in RuO$_2$ dominates the ordinary Hall effect contribution and the magnitude is comparable to metallic ferromagnets or noncollinear antiferromagnets~\cite{nagaosa2010anomalous,vsmejkal2022anomalous}.
Other metallic collinear altermagnets which are predicted or observed to demonstrate the crystal Hall effect include FeSb$_2$~\cite{mazin2021prediction, vsmejkal2022beyond}, VNb$_3$S$_6$~\cite{vsmejkal2022beyond} and CoNb$_3$S$_6$~\cite{vsmejkal2020crystal,ghimire2018large,tenasini2020giant,lu2022understanding}.

Studies of the AHE in spin-polarized antiferromagnets can provide insight into how magnetic orders beyond ferromagnetism can give rise to the Hall effect, and can thus pave the way for developing next-generation spintronics devices based on antiferromagnets.

\section{Nonlinear magnetotransport} \label{sec:nonlinear}
\begin{figure}[hbt!]
    \sidesubfloat[]{\includegraphics[width=0.3\linewidth]{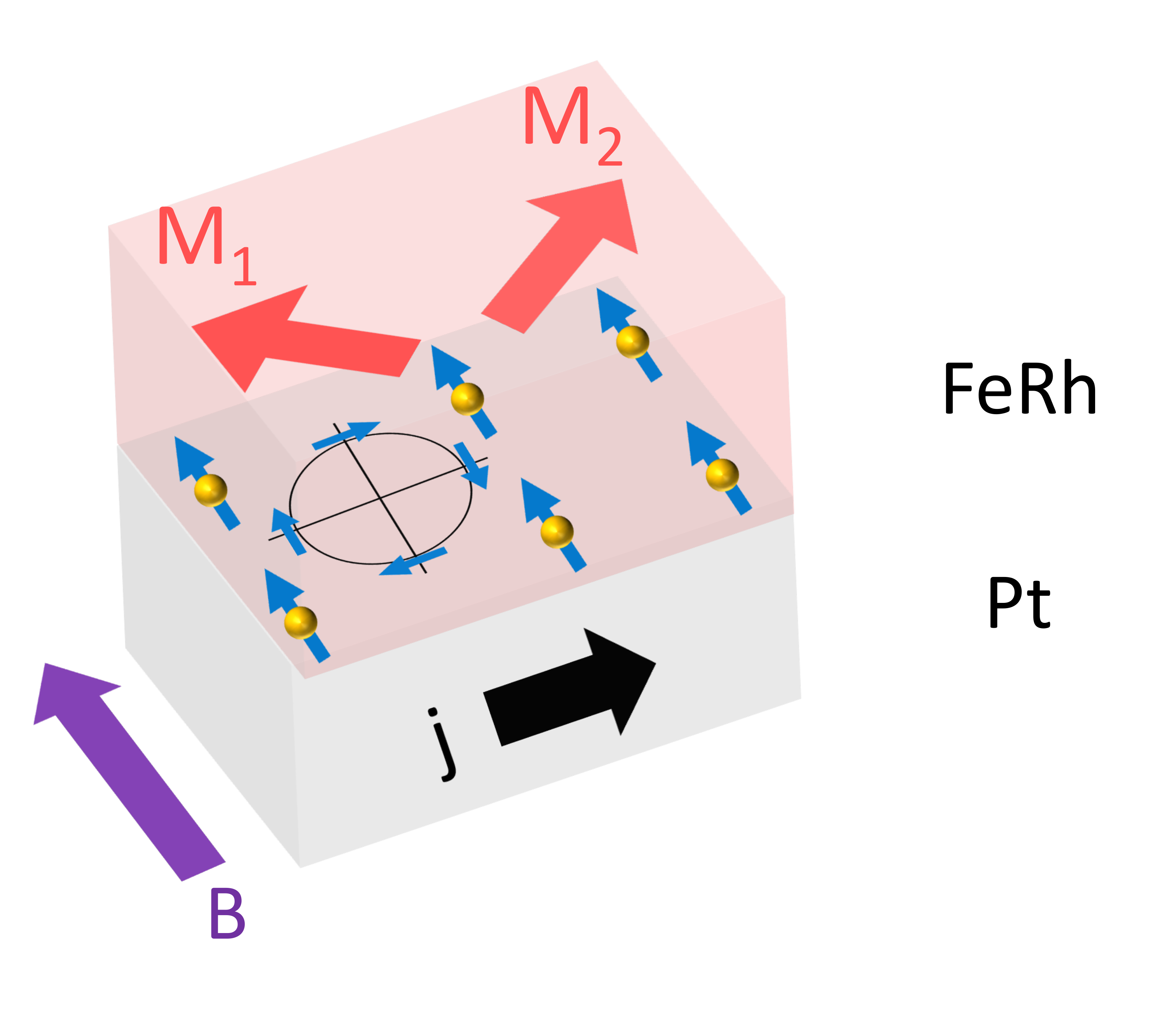}\label{fig:3_a}}\quad \quad%
    \sidesubfloat[]{\includegraphics[width=0.3\linewidth]{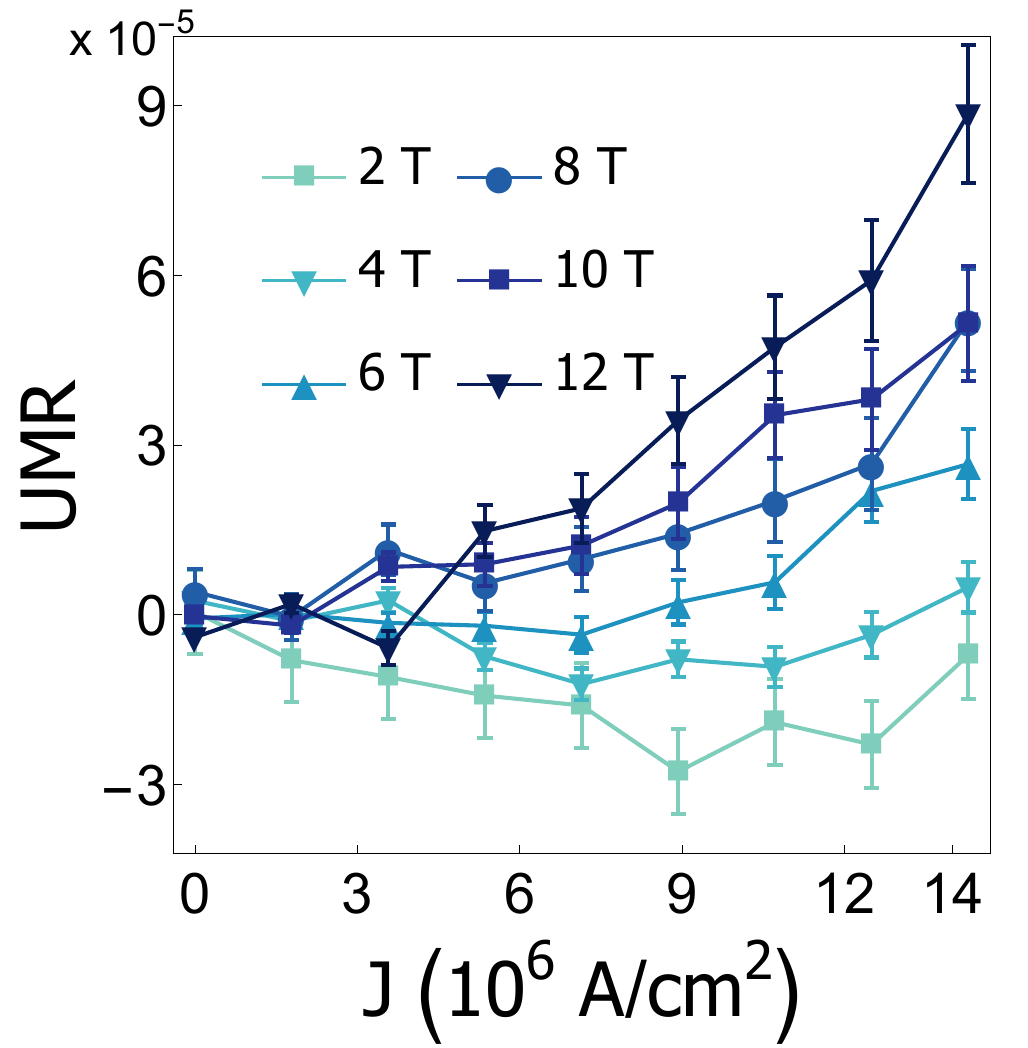}\label{fig:3_b}}\quad%
    \sidesubfloat[]{\includegraphics[width=0.25\linewidth]{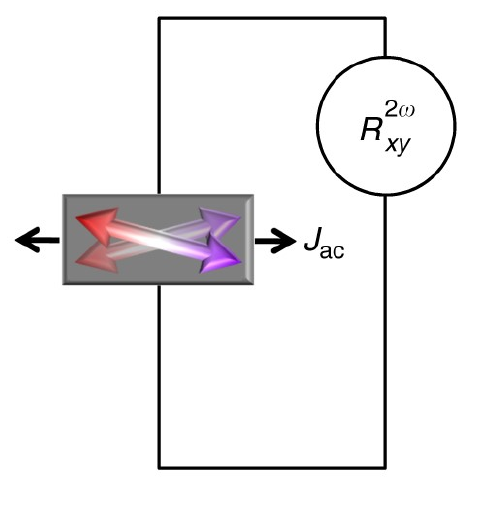}\label{fig:3_c}}\quad%
    \sidesubfloat[]{\includegraphics[width=0.3\linewidth]{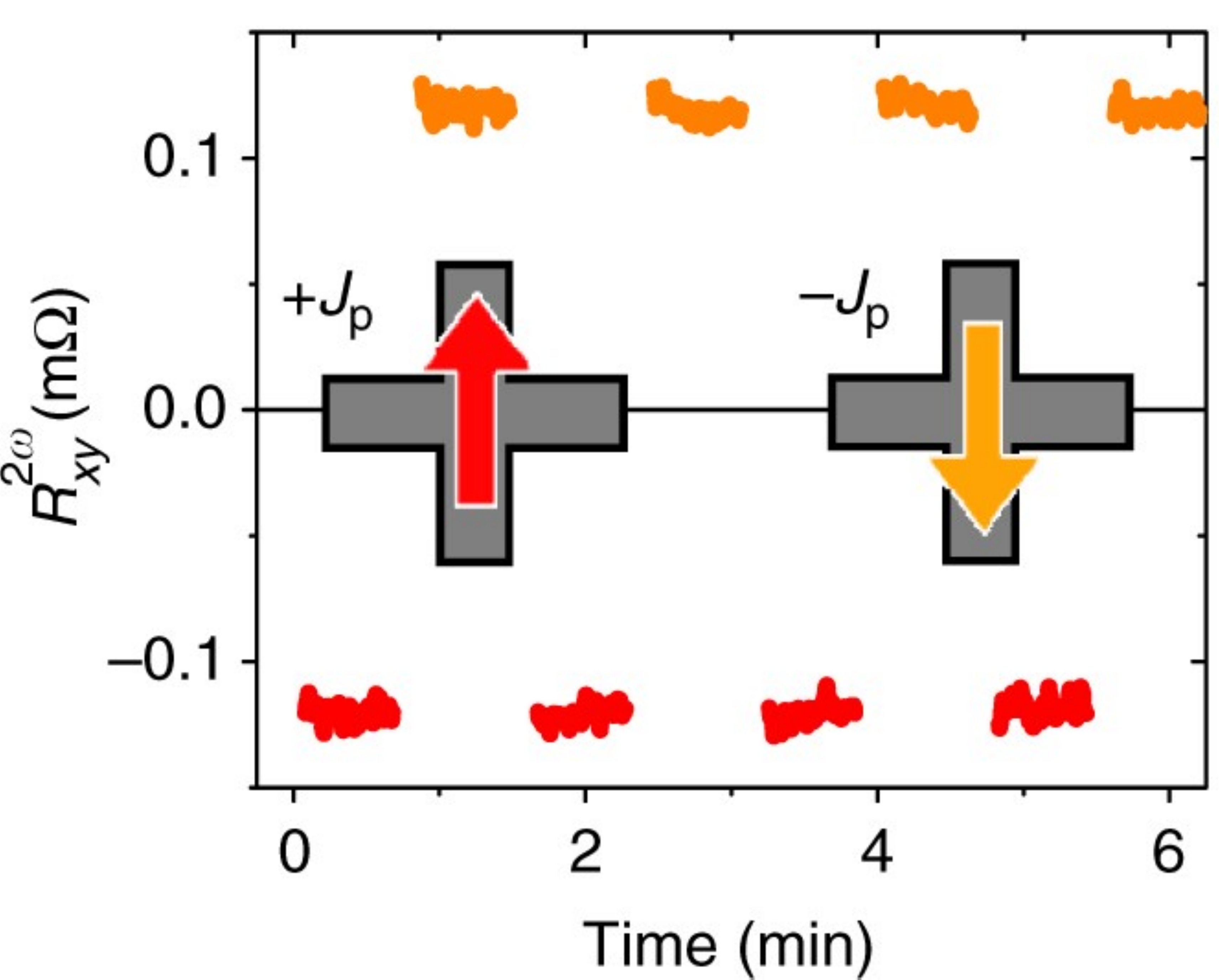}\label{fig:3_d}}%
    \caption{Nonlinear magnetotransport phenomena in spin-polarized antiferromagnets: \normalfont{(a,b) Unidirectional magnetoresistance (UMR) in FeRh/Pt bilayer. (a) Schematic of Rashba-effect induced spin-polarized electrons (blue arrows) and canted antiferromagnetic order (red arrows). (b) In-plane field dependence of UMR in antiferromagnetic FeRh/Pt bilayer. Panels (a) and (b) reprinted from Ref.~\cite{shim2022unidirectional} (CC 4.0 International). (c,d) Detection of Néel vector reversal in CuMnAs. (c) Measurement setup for characterizing the nonlinear Hall effect. The alternating probing current $J_{ac}$ is applied along x-axis and second-harmonic voltage is detected along y-axis ($R_{xy}^{2\omega}$). (d) Second-harmonic voltage can detect the 180\degree reversal of the Néel order. The writing current $J_{p}$ is applied along the y-direction to set the Néel vector and then the probing current is applied along the x-axis. Panels (c) and (d) reprinted from Ref.~\cite{godinho2018electrically} (CC 4.0 International).}}\label{fig:3}
\end{figure}

Nonlinear magnetotransport phenomena, namely unidirectional magnetoresistance (UMR) and nonlinear Hall effect, have recently attracted immense interest for their fundamental and practical implications. Nonlinear responses arise in systems with broken inversion symmetry and therefore expand the scope of magnetotransport beyond centrosymmetric systems. In addition, due to the unidirectional nature of nonlinear magnetotransport, nonlinear magnetotransport effects motivate novel detection schemes for antiferromagnetic order. In this section, we discuss how spin-polarized conduction electrons in metallic antiferromagnets enable new readout schemes of antiferromagnetic orders based on nonlinear magnetotransport effects, and allow the investigation of spin texture topology via the nonlinear Hall effect.

\subsection{Unidirectional magnetoresistance effects} \label{sec:nonlinear_UMR}
Just as linear magnetoresistance can be used to detect the spin state in antiferromagnetic systems, nonlinear magnetoresistance can also be utilized to probe magnetic order parameters. Unidirectional magnetoresistance (UMR) is an emerging nonlinear magnetoresistance effect whose sign changes when either the direction of the magnetization or that of the applied electric field is reversed~\cite{avci2015unidirectional,avci2018origins,liu2021magnonic,avci2015magnetoresistance}.  UMR can be hosted in systems with broken inversion symmetry and strong spin-orbit coupling, such as magnetic bilayers composed of a ferromagnetic material and a nonmagnetic material, wherein the effect can be attributed to the concerted action of current-induced spin polarization and spin-dependent electron scattering~\cite{olejnik2015electrical, zhang2016theory, yasuda2016large, avci2018origins, liu2021magnonic}. Its unidirectionality provides a potentially useful way of reading out the spin state in the ferromagnetic layer.

Recently, it was demonstrated that the UMR effect can also arise in an antiferromagnetic FeRh/Pt bilayer when the Néel vector cants with an applied field~\cite{shim2022unidirectional} (Figs.~\ref{fig:3_a} and ~\ref{fig:3_b}). The observed UMR exhibits a strong magnetic field dependence up to 12 T and undergoes a sign change even when the direction of the applied magnetic field is fixed (see Fig.~\ref{fig:3_a}), at variance with the UMR observed in ferromagnetic bilayer systems~\cite{avci2018origins}. Physically, the UMR may be attributed to the interplay between canted antiferromagnetism and current-induced spin-polarization in the interfacial Rashba states.   The field-induced canting in FeRh conspires with the large antiferromagnetic exchange coupling to induce a strong effective magnetic field, which significantly distorts the band structure and gives rise to the sizable UMR. The role of spin canting in inducing UMR in antiferromagnetic systems has been further confirmed by studies in canted antiferromagnetic-insulator/Pt bilayers~\cite{cheng2023unidirectional,fan2022observation,zheng2023coexistence}.

While UMR has so far only been observed in antiferromagnet/heavy-metal bilayer systems with field-induced spin canting, nonlinear magnetoresistance effects may also be hosted in intrinsically spin-polarized antiferromagnets, such as noncollinear antiferromagnets and collinear altermagnets. Exploration of nonlinear magnetoresistance effects in spin-polarized antiferromagnets may provide deeper insights into the scattering mechanisms between spin-polarized electrons and the local moments in antiferromagnets and altermagnets and may, in turn, pave the way for developing effective means for the detection and control of their magnetic order parameters for future spintronics devices.

\subsection{Nonlinear Hall effects} \label{sec:nonlinear_NLHE}    
Similar to how the anomalous Hall effect in linear response has been a macroscopic probe of the spin-orbit coupling and band topology in magnetic systems, nonlinear Hall effects may also serve as a powerful tool in probing important quantum geometric properties of Bloch states in emerging magnetic materials, especially for those lacking inversion symmetry. Formally, information about the quantum geometry is encoded in the quantum geometric tensor, which has both an imaginary part (i.e., the Berry curvature) and a real part (i.e., the quantum metric)~\cite{provost1980riemannian}. While the intrinsic anomalous Hall effect originates from the Berry curvature monopole, nonlinear responses may emanate from the dipole and multipole moments of either the Berry curvature or quantum metric~\cite{gao2014field, sodemann2015quantum, wang2023quantum, gao2023quantum}.



Recent reports have shown that the nonlinear Hall effect can probe both Berry curvature and quantum metric phenomena. For instance, a sizable nonlinear Hall effect was observed in CuMnAs~\cite{godinho2018electrically}---a collinear antiferromagnet with $\mathcal{P}$ and $\mathcal{T}$ symmetries separately broken but the combined $\mathcal{PT}$ symmetry being preserved---which may be attributed to the intrinsic contribution from the dipole moment of the Berry-connection polarizability tensor~\cite{wang2021intrinsic,liu2021intrinsic}. The nonlinear Hall effect is also deemed to be potentially useful, as it can be utilized to electrically distinguish opposite Néel vector orientations (Figs.~\ref{fig:3_c} and ~\ref{fig:3_d}). As illustrated in Fig.~\ref{fig:3_c}, due to the broken $\mathcal{P}$ symmetry in CuMnAs, electrical current generates staggered non-equilibrium spin polarization, which causes transient deflection of the antiferromagnetic order (thick red and purple arrows on Mn sites). Since the Néel vector deflects in opposite directions depending on its orientation, it is possible to detect its 180\degree reversal by measuring the second-harmonic transverse resistance $R_{xy}^{2\omega}$ (Fig.~\ref{fig:3_d}). Other recent works have reported the nonlinear Hall effect in even-layered MnBi$_2$Te$_4$ (a collinear $\mathcal{PT}$-symmetric antiferromagnetic system), which originates from the quantum metric dipole~\cite{wang2023quantum,gao2023quantum}.

While recent studies of nonlinear Hall effects were mostly concentrated on $\mathcal{PT}$-symmetric antiferromagnets with spin-degenerate bands, spin-polarized antiferromagnets with broken $\mathcal{PT}$ symmetry are expected to provide another exciting materials platform for exploring nonlinear Hall effects with different physical origins. A nonlinear anomalous Hall effect has been predicted to emerge in a half-Heusler alloy CuMnSb (an antiferromagnetic metal with broken $\mathcal{PT}$ symmetry), wherein the combination of magnetic crystal group symmetries and spin-orbit coupling induces a large Berry curvature dipole~\cite{shao2020nonlinear}. Nonlinear Hall effects associated with higher order moments of Berry curvature and quantum metric in altermagnets have been also predicted~\cite{fang2023quantum}. Experimental observation of such nonlinear Hall effects as well as the discrimination between them when they coexist may greatly deepen our understanding of the quantum geometry in these emerging magnetic materials and shed light on pathways to engineer the geometric properties of electronic states for future device applications. 

In addition to the aforementioned intrinsic mechanisms rooted in quantum geometry, the interplay between spin-split bands and spin-dependent scatterings may also come into play in magnetotransport, giving rise to extrinsic nonlinear Hall effects. For instance, the nonlinear planar Hall effect in canted antiferromagnets---a counterpart of the UMR effects therein~\cite{shim2022unidirectional}---has yet been detected. The observation of such an effect will provide valuable insights into how itinerant electrons occupying momentum-dependent spin-split bands are scattered against various scattering sources (e.g., impurities, magnons, etc.) in the nonlinear transport regime.

\section{Spin current generation} \label{sec:spin current}

\begin{figure}[hbt!]
    \sidesubfloat[]{\includegraphics[width=0.5\linewidth]{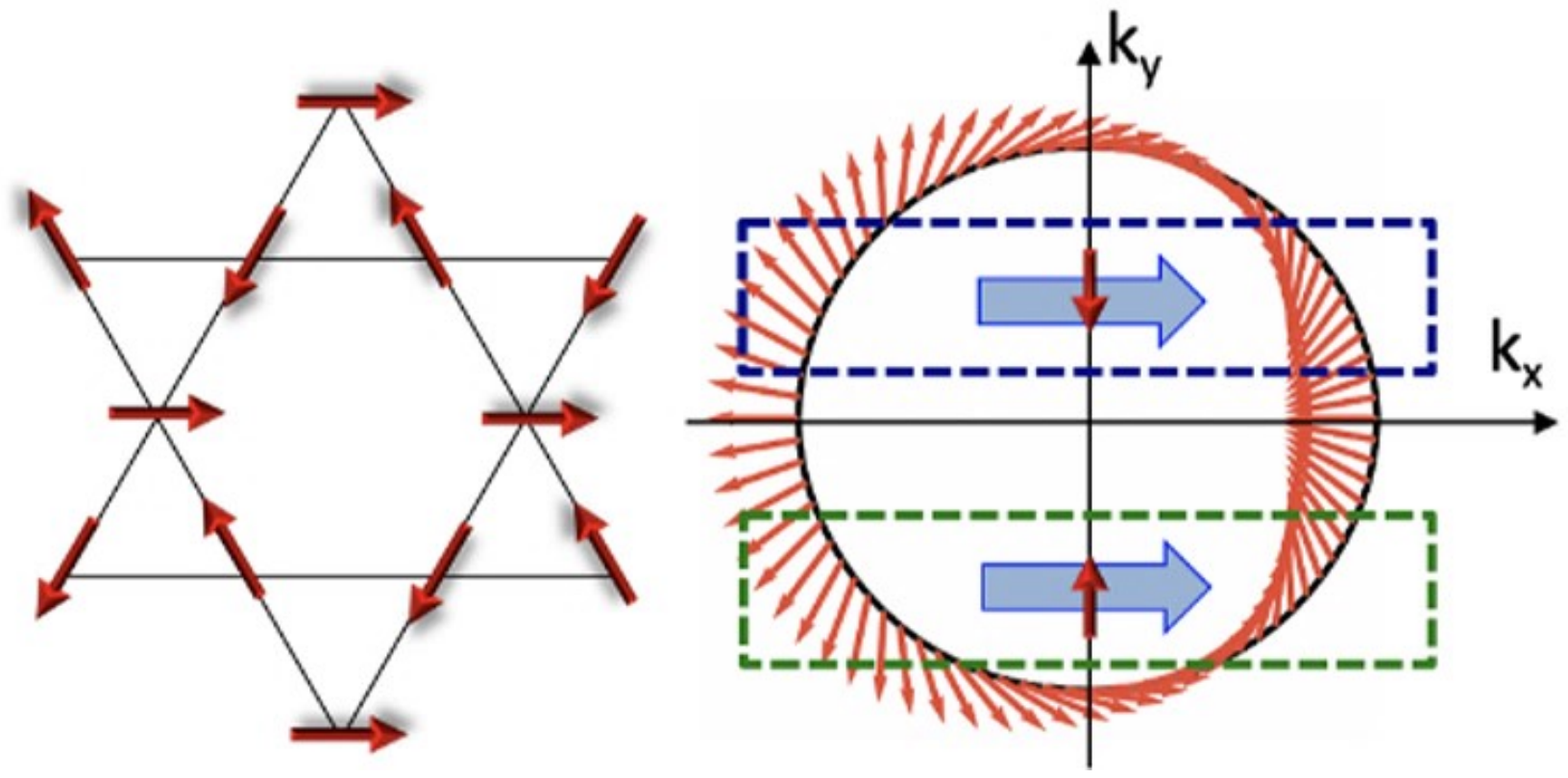}\label{fig:4_a}}\quad%
    \par \bigskip \bigskip
    \sidesubfloat[]{\includegraphics[width=0.35\linewidth]{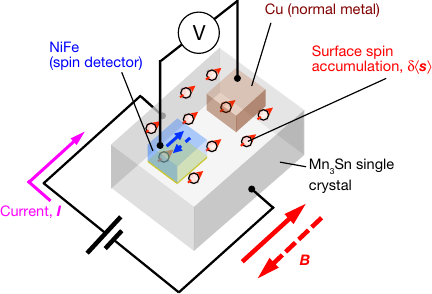}\label{fig:4_b}}\quad%
    \sidesubfloat[]{\includegraphics[width=0.55\linewidth]{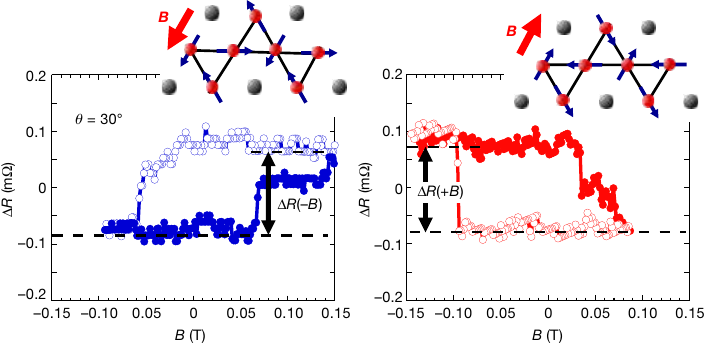}\label{fig:4_c}}%
    \caption{Magnetic spin Hall effect in spin-polarized antiferromagnets: \normalfont{(a) Schematics explaining the origin of the magnetic spin Hall effect in a noncollinear antiferromagnets on a kagomé lattice. Magnetic configuration (left) and the corresponding spin texture (right) in momentum space at the Fermi level. Non-equilibrium magnetic spin Hall current is spin polarized along y-direction. Panel (a) reprinted with permission from Ref.~\cite{bonbien2021topological}. Copyright IOP Publishing Ltd. All rights reserved. (b,c) Magnetic spin Hall effect observed in Mn$_3$Sn. (b) Non-local spin-valve device setup to measure the magnetic spin Hall effect (c) A non-local magnetoresistive signal demonstrating the existence of non-equilibrium spin accumulation whose symmetries are consistent with the magnetic spin Hall effect. Panels (b) and (c) reprinted with permission from Ref.~\cite{kimata2019magnetic}, Springer Nature.}}\label{fig:4}
\end{figure}

In this section, we discuss how spin-polarized antiferromagnetic metals can generate spin currents based on their spin-polarized conduction electrons. Generating spin-polarized current is a central element in most spintronics applications. Recently, it has been demonstrated that the spin Hall effect, in which charge current induces a transverse spin accumulation due to spin-orbit interactions~\cite{dyakonov1971possibility,hirsch1999spin,zhang2000spin}, can generate a pure spin current and the spin Hall effect can exist in a wide range of materials: semiconductors~\cite{kato2004observation,wunderlich2005experimental}, metals~\cite{saitoh2006conversion,valenzuela2006direct,kimura2007room}, superconductors~\cite{wakamura2015quasiparticle}, ferromagnets~\cite{miao2013inverse}, and also antiferromagnets~\cite{mendes2014large,zhang2014spin,zhang2015all}.

Early studies of the spin Hall effect in metallic antiferromagnets were focused on the conventional spin Hall effect originating from Berry curvature induced by spin-orbit coupling. Sizable spin Hall effects have been reported in metallic antiferromagnets such as sputtered Ir$_{20}$Mn$_{80}$~\cite{mendes2014large} and CuAu-I-type antiferromagnets, FeMn, PdMn, IrMn, and PtMn \cite{zhang2014spin,zhang2015all}. In particular, for MnX compounds (X = Ir, Pt, Pd, Fe), it was reported that the spin Hall angle increases with the atomic weight of the heavy metal alloyed with Mn, and it becomes comparable to Pt \cite{zhang2014spin}. Such an increase in the spin Hall angle was attributed to the heavy-metal element in these alloys and the modification of the band structure by the magnetic elements. In addition to the increase in the spin Hall angle, it has been found that spin Hall effects depend strongly on the orientation of the antiferromagnetic moments \cite{zhang2015all}. The conventional antiferromagnet serving as a spin source is also reported in the antiferromagnetic metal FeRh, in which the antiferromagnetic ordering in FeRh enables exotic spin torques~\cite{gibbons2022large}.

Recently, it has been predicted~\cite{vzelezny2017spin,mook2020origin} and experimentally observed~\cite{kimata2019magnetic,holanda2020magnetic} that the magnetic ordering in metallic antiferromagnets can generate spin currents with unconventional spin-polarization directions due to symmetry breaking. Such a spin Hall effect due to the magnetic ordering is called a magnetic spin Hall effect, and it is reminiscent of conventional spin Hall effects in CuAu-I type metallic antiferromagnets, which are highly anisotropic with the orientation of the antiferromagnetic moments~\cite{zhang2015all}.

Fig.~\ref{fig:4_a} shows how the magnetic spin Hall effect can arise in a noncollinear antiferromagnet on a kagomé lattice. Due to the anisotropic spin texture in a noncollinear antiferromagnet, a charge current flowing in the x direction will induce different net spin polarizations for states with $k_y > 0$ (blue box in the spin texture of Fig.~\ref{fig:4_a}) and $k_y < 0$ (green box in the spin texture of Fig.~\ref{fig:4_a}). Thus, unlike the conventional spin Hall effect of which spin polarization is governed only by the charge current and pure spin current directions, for the magnetic spin Hall effect, the spin polarization of the generated spin current depends on the magnetic texture in momentum space and the specific magnetic configuration of the antiferromagnet.

The magnetic spin Hall effect has been observed in the noncollinear antiferromagnetic metal Mn$_3$Sn. As illustrated in Fig.~\ref{fig:4_b}, in a non-local magnetic spin-valve, the spin accumulation generated by the spin Hall effect was measured at the interface between Mn$_3$Sn and NiFe. An observed magnetoresistive signal switches its sign depending on the magnetic configuration of the antiferromagnet and such correlation suggests that the spin accumulation likely originates from magnetic spin Hall effect (Fig.~\ref{fig:4_c}). Another recent work reported a large current-induced modulation of the magnetic damping of a Ni$_{80}$Fe$_{20}$ (Permalloy)/Mn$_3$Ir bilayer which strongly depends on the crystal orientation of Mn$_3$Ir~\cite{holanda2020magnetic}. Such modulation was also attributed to the magnetic spin Hall effect in the noncollinear antiferromagnet Mn$_3$Ir.

The magnetic spin Hall effect has also been extended to collinear altermagnetic metals. Following the theoretical prediction~\cite{gonzalez2021efficient}, it was experimentally demonstrated that RuO$_2$ can generate a spin current in which the spin current direction is correlated to the crystal orientation of RuO$_2$ and a spin polarization is aligned approximately with the Néel vector~\cite{bose2022tilted,bai2022observation}. The observed orientation of spin polarization was consistent with the magnetic spin Hall effect and subsequent experimental works provide insights into how the crystal axes~\cite{bai2023efficient} or antiferromagnetic ordering ({\em e.g.}, 
incommensurate spin density wave)~\cite{feng2024incommensurate} of RuO$_2$ affects the magnetic spin Hall effect in altermagnets.

Studies of the magnetic spin Hall effect in spin-polarized antiferromagnetic metals can provide insight into the role of spin textures in spin current generation; such insights can pave the way for developing tunable spintronics devices based on magnetic orders.

\section{Spin torques} \label{sec:spintorques}

In this section, we discuss how spin-polarized antiferromagnetic metals can generate spin torques based on their spin-polarized conduction electrons. Spin currents can apply torques to magnetic moments by transferring angular momentum from spin-polarized carriers to local magnetic moments. In ferromagnetic spintronics research, spin torques generated by antiferromagnets have been used to excite magnetization dynamics in adjacent ferromagnets.  Over time, the literature has shifted focus on the role that the antiferromagnetic order itself plays in generating spin torques.   Using the spin torque ferromagnetic resonance (ST-FMR) technique, spin torques originating from conventional spin Hall effects in metallic antiferromagnets FeMn, PdMn, IrMn, and PtMn~\cite{zhang2015all} were successfully used to excite magnetization dynamics in adjacent ferromagnets.  Similar experiments, using the antiferromagnetic spin Hall effect of Mn$_2$Au~\cite{chen2021observation} demonstrated both the excitation of magnetization dynamics as well as the field-free switching of the magnetization of an adjacent ferromagnet, providing evidence for an out-of-plane polarized spin current.  In FeRh, temperature dependent ST-FMR demonstrated that the effect of spin torques on adjacent ferromagnetic materials was sensitive to the magnetic ordering (ferro- or antiferro-) of the material~\cite{gibbons2022large}.  More recently, in RuO$_2$/Permalloy heterostructures, ST-FMR unveiled unconventional damping-like spin torques.  Here, the in-plane and out-of-plane spin polarization of the damping-like torque was found to be dependent on both the crystal orientation of the RuO$_2$ as well as the Néel vector orientation~\cite{karube2021observation, bose2022tilted, bai2022observation}.

Beyond being used as materials capable of generating spin torques, there has also been a significant focus on the role of spin torques acting on antiferromagnetic order.  This is largely because antiferromagnets have antiferromagnetic resonance (AFMR) and long wavelength magnon modes with frequencies that span tens~\cite{macneill2019gigahertz}, hundreds~\cite{vaidya2020subterahertz,li2020spin}, and thousands of GHz~\cite{moriyama2019intrinsic}.  The ultrafast magnetization dynamics found in these materials make antiferromagnetic materials attractive for next generation antiferromagnetic spintronic technologies~\cite{baltz2018antiferromagnetic}, such as THz spin torque oscillators and antiferromagnetic magnetic memory technologies~\cite{marti2014room}. There has also been experimental progress in probing the ultrafast magnetization dynamics found in antiferromagnets with spintronic experimental methods.  For example, spin pumping from sub-THz excited antiferromagnetic resonance has been reported in MnF$_2$~\cite{vaidya2020subterahertz} and Cr$_2$O$_3$~\cite{li2020spin}.  More recently, the spin torque driven antiferromagnetic resonance of Fe$_2$O$_3$ has also been reported~\cite{zhou2024spin}.  However, these types of experiments, which emulate experiments performed with ferromagnets, are generally limited to antiferromagnetic insulator materials.  Experimental studies involving ultrafast magnetization dynamics in spin-polarized antiferromagnetic metals are still generally lacking.  There has been theoretical progress regarding the effects of spin pumping into an altermagnet via ferromagnetic resonance of an adjacent ferromagnetic insulator.  Here it was shown that the spin current pumped into the altermagnet can be both enhanced and suppressed due to the crystallographic orientation of the altermagnet at the interface with the ferromagnet~\cite{sun2023spin}.  Interconnections and implications between this anisotropic spin pumping effect and spin torques that could be generated via spin injection into altermagnets have not yet been explored further.

One of the great successes that has taken place in antiferromagnetic metals has been the electrical ``switching'' or reorientation of the Néel vector.  In certain antiferromagnetic metals, such as CuMnAs and Mn$_2$Au~\cite{wadley2016electrical,bodnar2018writing}, the passing of an electric current through the material induces staggered relativistic spin-orbit torques on the individual magnetic sub-lattices of the antiferromagnet. Such torques can change the Néel vector orientation averaged across various domains within the material~\cite{grzybowski2017imaging}.  The change in the Néel vector orientations within the sample can be ``read'' out via the AMR of the antiferromagnet itself.   In the non-collinear antiferromagnet, Mn$_3$Sn, spin-orbit torques generated from adjacent Pt thin films were used to reorient the antiferromagnetic order~\cite{takeuchi2021chiral}.  Here, the Hall resistivity was used to read out the state of the antiferromagnet after current pulses of various amplitudes were passed through the heterostructure.    More recently, it has been theoretically predicted that in the altermagnet RuO$_2$ and in Fe$_4$GeTe$_2$ globally spin neutral currents can be decomposed into ``Néel'' spin currents~\cite{shao2023neel}.  Néel spin currents can be thought of as staggered, non-neutral, spin currents flowing in separate magnetic sub-lattices within an antiferromagnet.  Theoretically, it was shown that in antiferromagnetic tunnel junctions comprised of such materials, spin torques generated by such Néel spin currents could be used to electrically switch the Néel vector within such a tunnel junction.

Already, large and unconventional spin torques generated by antiferromagnets, and exerted on ferromagnets, are being used to create new opportunities for the excitation and switching of ferromagnetic magnetization.  Looking ahead, the experimental demonstration of driven antiferromagnetic magnetization dynamics and the switching of Néel vectors with such unconventional spin torques remains an open opportunity for the broader spintronics community to explore.

\section{Conclusion and Outlook} \label{sec:conclusion}

Spin-polarized antiferromagnetic metals have recently gained increasing interest due to the direct and strong coupling between charge transport and magnetic spin textures in antiferromagnetic metals. As recent advances suggest, spin textures in metallic antiferromagnets can host new types of charge transport phenomena including linear and nonlinear magnetotransport, spin current generation, and spin torque generation.

The investigation of spin-polarized antiferromagnetic metals is in the early stage and many future research directions are left to be explored. 
One possible research direction for spin-polarized antiferromagnetic metals is to explore the interplay between spin-polarized antiferromagnetic order and superconductivity. It has been demonstrated that the Berry curvature of the noncollinear antiferromagnetic metal Mn$_3$Ge can induce a long-range supercurrent in superconductor/Mn$_3$Ge/superconductor lateral Josephson junctions~\cite{jeon2021long}. Such a long-range supercurrent is attributed to the noncollinear spin structure of Mn$_3$Ge allowing the spin-mixing and spin-rotation mechanisms required for singlet-to-triplet pair conversion, and thus enabling the long-range spin-triplet supercurrents. Apart from noncollinear antiferromagnets, several theoretical works suggest that canted antiferromagnets can host spin-triplet Cooper pairs~\cite{chourasia2023generation} and altermagnets may also be compatible with both spin-singlet and spin-triplet Cooper pairing mechanisms due to their anisotropic Berry curvatures~\cite{vsmejkal2022emerging}. Moreover, Josephson junctions based on altermagnets are predicted to have an added advantage over ferromagnetic Josephson junctions in that their supercurrents can be tuned with the crystalline orientation of the altermagnet~\cite{ouassou2023dc,zhang2024finite}. The predicted connection between altermagnets and unconventional superconductivity suggests that spin-polarized antiferromagnetic metals hold promise in the context of unconventional superconductivity.

The identification and investigation of material systems with spin-polarized antiferromagnetism is also in the early stages. Spin-polarized antiferromagnets, such as altermagnets, can exist in two-dimensional (2D) and three-dimensional (3D) crystals~\cite{vsmejkal2022emerging}. Thus, beyond widely studied noncollinear antiferromagnets Mn$_3$X family or the collinear altermagnet RuO$_2$, there may be multiple spin-polarized antiferromagnetic metals that are yet unexplored but expected to host new types of emergent phenomena. For instance, integrating 2D spin-polarized antiferromagnetic metals such as VNb$_3$S$_6$~\cite{vsmejkal2022beyond} and CoNb$_3$S$_6$~\cite{vsmejkal2020crystal} with other van der Waals materials may open up opportunities to explore the interplay between spin, charge, and valley degrees of freedom with great tunability. Studying different metallic material platforms that can interface spin-polarized antiferromagnetic order with other degrees of freedom awaits further experimental investigations and will provide valuable insights into fundamental physical concepts and the development of novel spintronics devices.

\section{Acknowledgements}
This work was primarily supported by the NSF through the University of Illinois Urbana−Champaign Materials Research Science and Engineering Center under Awards DMR-1720633 and DMR-2309037. J.S. acknowledges support from the National Science Foundation under DMR-2328787. Work by S. S.-L. Z and M. M. was partly supported by the National Science Foundation (Award No. 2326528).

\bibliographystyle{ar-style4}
\bibliography{mainref}
\end{document}